\documentclass[preprint2]{aastex}

\begin{document}

\title{Nucleosynthesis in the Outflow from Gamma Ray Burst 
Accretion Disks}

\author{R. Surman\altaffilmark{1}, G. C. McLaughlin\altaffilmark{2} and W. R. Hix\altaffilmark{3}}

\altaffiltext{1}{Department of Physics and Astronomy, Union College, Schenectady, NY 
12308}
\altaffiltext{2}{Department of Physics, North Carolina State University, 
Raleigh, NC  27695-8202}
\altaffiltext{3}
{Physics Division, Oak Ridge National Laboratory, Oak Ridge, TN 37831-6374}

\begin{abstract} 

We examine the nucleosynthesis products that are produced in the outflow
from rapidly accreting disks.  We find that the type of element synthesis
varies dramatically with the degree of neutrino trapping in the disk and
therefore the accretion rate of the disk.  Disks with relatively high
accretion rates such as $\dot{M} = 10 \, M_\odot/ {\rm s}$ can produce
very neutron rich nuclei that are found in the $r$ process.  Disks with
more moderate accretion rates can produce copious amounts of Nickel as
well as the light elements such as Lithium and Boron.  Disks with lower
accretion rates such as $\dot{M} = 1 \, M_\odot/ {\rm s}$ produce large
amounts of Nickel as well as some unusual nuclei such as $^{49}{\rm Ti}$,
$^{45}{\rm Sc}$, $^{64}{\rm Zn}$ and $^{92}{\rm Mo}$.  This wide array of
potential nucleosynthesis products is due to the varying influence of
electron neutrinos and antineutrinos emitted from the disk on the
neutron-to-proton ratio in the outflow. We use a parameterization for the
outflow and discuss our results in terms of entropy and outflow
acceleration. 

\end{abstract}


\section{Introduction}

Data on long duration gamma ray bursts taken over the last few years has
sparked the idea that these events are associated with core collapse
supernovae, for a review see \citet{mes02}.  Theoretical models, e.g. the
collapsar model \citep{woo93,mac99}, suggest that a massive star collapses
to an accretion disk surrounding a black hole, and a jet is driven out of
the center of the star, leaving behind an escape route for the burst
itself.  Such an event will produce nucleosynthesis, not just in the the
jet, but in the slower outflow that comes off of the accretion disk, as
well as through explosive burning. 

It is not known whether there are two types of events, core collapses that
make ``normal'' supernovae with proto-neutron star cores and those that
have accretion disks surrounding black holes that produce gamma ray
bursts, or whether there is there a continuum of objects in between these
extremes. The answer to the question, while of interest in its own right,
is of crucial importance as far as our understanding of the significance
of these objects for element synthesis.  Estimates of the rate of gamma
ray bursts from optical observations show that they occur on a timescale
of less than about $10^{-2}$ times the rate of core collapse supernovae
\citep{pod04}, while radio observations constrain the rate to be less than
5\% of the Type Ib/c rate \citep{berger03}. However, accretion disks which
form from stellar collapse may be a more common phenomenon, and and
consequently element synthesis in outflow from an accretion disk may be
much more frequent. 

We have less observation evidence for the origin of short duration gamma
ray bursts. Only one afterglow of a short (30 ms) gamma-ray burst has been
detected thus far (GRB 050509b).  Theoretical calculations have shown that
these events may plausibly occur as a result of neutron star mergers, e.g.
\citet{ruf99,ruf01,ros03a,ros03b}.  Numerical simulations of these events
have produced very rapidly accreting disks which surround black holes with
rates between $\dot{M} =1 \, M_\odot / {\rm s}$ and $\dot{M} = 10 M_\odot
/ {\rm s}$. 

We present a study of the nucleosynthesis products produced from the
outflow from accretion disks of $\dot{M} =0.1 \, M_\odot / {\rm s}$,
$\dot{M} =1 \, M_\odot / {\rm s}$, and $\dot{M} = 10 M_\odot / {\rm s}$
which have black hole spin parameters of $a=0$.  In general neutrino
trapping, as a consequence of higher temperatures, will occur for lower
accretion rate disks if the spin parameter is larger.  We show an example
of the element synthesis from an $a=0.95$, $\dot{M} =0.1 \, M_\odot / {\rm
s}$ disk to illustrate the effect of higher spin. Nucleosynthesis in gamma
ray burst disk winds has also been studied by \citet{pru03,fuji03,fuji04}. 

Neutrinos can become trapped in the disks, particularly those that are
associated with neutron star mergers.  The emerging neutrinos annihilate
and may drive the burst itself.  Therefore neutrino trapping in these
disks is under study by several groups, e.g.
\citet{lee,ros03b,kohri,januik,set04}.

The neutrinos have an enormous influence on the final elemental
abundances, even in cases where the neutrinos are not trapped in the disk.
The disks are hot enough to release copious numbers of electron neutrinos
and antineutrinos that then interact with free neutrons and protons in the
outflow from the disk.  Their influence over the neutron-to-proton ratio,
together with entropy, and therefore heating of the material which leaves
the disk, determines the final nucleosynthesis products. 

We pay particular attention to the abundance of Nickel-56.  Nickel-56
drives the light curves of core collapse supernovae.  Since supernova
light curves have been detected in gamma ray burst light curves, this
implies substantial synthesis of Nickel-56 either in the outflow from the
disk \citep{pru04} or in explosive burning \citep{maeda03}.  The amount of
Nickel observed in the light curves corresponds to roughly half of a solar
mass \citep{woo03}. 

We also examine some nuclei whose origin is poorly understood, but have
large overproduction factors in the outflows from low accretion rate
disks, such as Titanium-44, Scandium-45 and Zinc-64 \citep{psm04}.  We
look at the light $p$-process nuclei such as $^{92}{\rm Mo}$ and
$^{94}{\rm Mo}$.  Rapidly accreting disks have been suggested as a site
for the production of such light $p$-nuclei in \citet{fuji03}. 

We examine the prospects for producing an $r$ process in high accretion
rate disks by way of material which is driven neutron rich by neutrino
interactions as discussed in \citet{ms05}.  And we also discuss conditions
under which small amounts of very light nuclei such as Lithium and Boron
can be produced in medium accretion rate disks. 

\section{Disk and Outflow models}

We use disk models from \citet{pop99} (PWF) for the low accretion rate
disk, and from \citet{dim02} (DPN) for the high accretion rate disk.  The
PWF models do not include neutrino trapping and indeed for accretion rates
less than $ \dot{M} = 0.1 \, M_\odot / {\rm s}$ there is very little 
trapping.  For higher accretion rates neutrinos do become trapped and
therefore we use the DPN models which include effects of neutrino
trapping. 

For the outflow we use the trajectories calculated in \citet{sur05} and 
parameterize the outflow as a wind using
\begin{equation}
|u| = v_\infty \left(1 - {R_0 \over R} \right)^\beta
\label{eq:velocity}
\end{equation}
where the starting position of the material is $R_0$ and $\beta$
determines the acceleration of the wind.  Large beta represents slow
acceleration whereas small beta represents rapid acceleration. 

In the disk the entropy in units of Boltzmann's constant is $s/k \sim 10$. 
Since the material may be heated as it leaves the disk, we consider a 
range of entropies between  $s/k =10$ and $s/k = 50$.

We use the neutrino fluxes calculated in \citet{sur04} for each of the the
disk models we use here.  In order to calculate the neutron-to-proton
ratio in the outflowing trajectory we integrate of the flux coming from
all parts of the disk, as described in \citet{sur04}. 

The place where the outflow occurs from the disk will have an impact on the
nucleosynthesis, primarily because the neutrino flux varies over the length of
the disk.  In our calculations we use starting points of $r_{0}=100$ km and 
$r_{0}=250$ km from the black hole.

We begin the elemental evolution in the disk outflow by dynamically
following the evolution of the electron fraction in the outflow, as
described in \citet{sur05}.  As the temperature drops below $10^{10}$ K,
we follow nuclear recombination using an intermediate network calculation
\citep{raph} that includes strong, electromagnetic, and weak interactions. 
Finally we use a network that does solely neutron capture,
photo-disintegration, beta decay, charged-current neutrino interactions,
and beta-delayed neutron emission \citep{rs}.  The latter is necessary for
calculations of the most neutron rich elements. 

\section{Nucleosynthesis from disks with Low Accretion Rates} \label{low}

Using the outflow trajectories described in the previous section we
present the results of reaction network calculations for outflow from a
PWF disk with an accretion rate of $\dot{M} = 0.1 \, M_\odot \, / {\rm
s}$.  This is representative of the accretion rates produced by the
collapsar model \citep{mac99}. 

\begin{figure}
\plotone{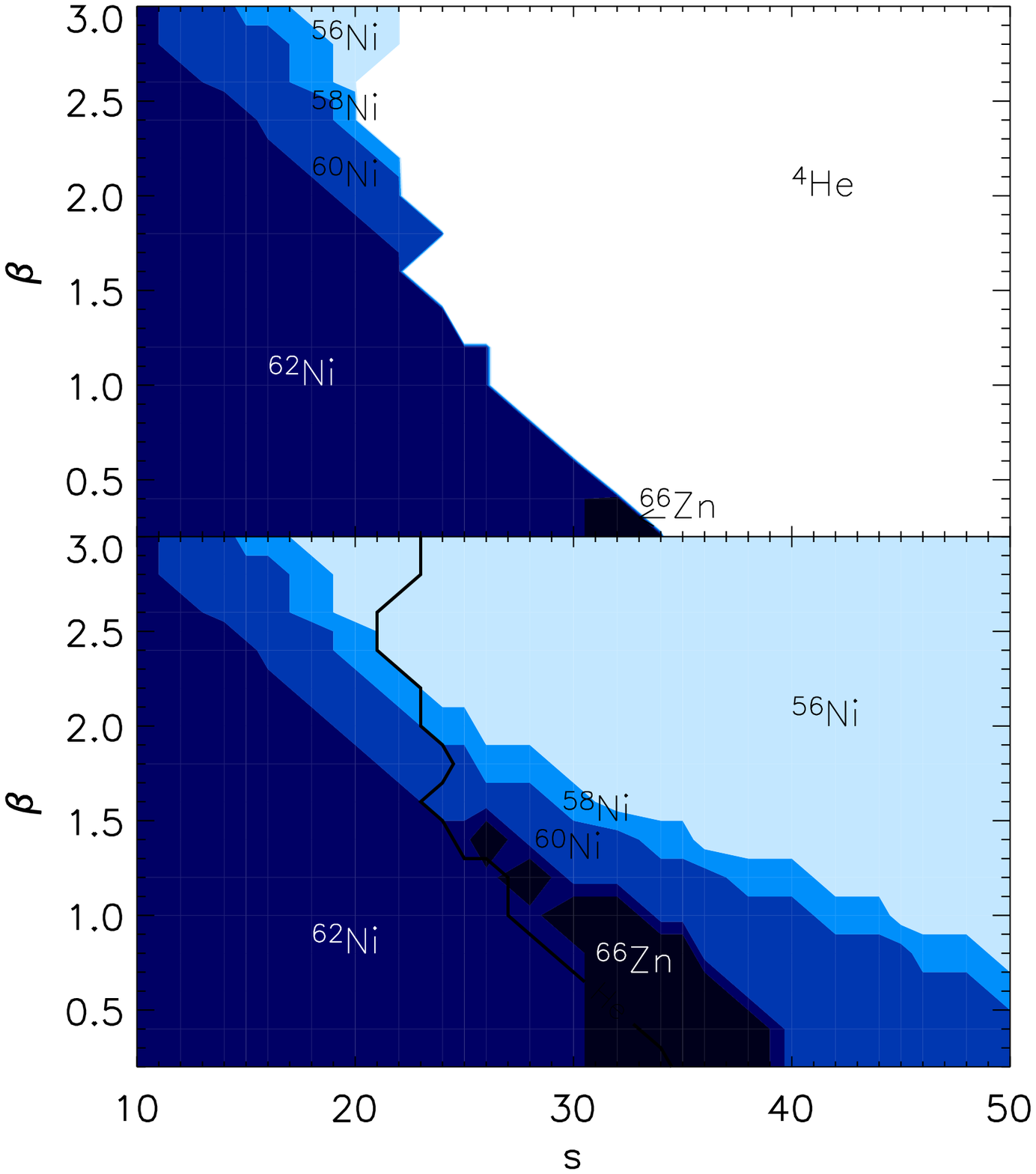}
\caption{Shows the nuclei with the largest mass fraction in the
outflow from a low accretion rate disk: $\dot{M} = 0.1 \, 
{\rm M}_\odot / {\rm s}$. The vertical and horizontal axes are entropy 
per baryon, $s$, and the acceleration parameter, $\beta$.
The lower panel shows the isotope with the largest mass fraction, 
excluding Helium-4.
\label{fig:xmass_low}}
\end{figure}

In Fig \ref{fig:xmass_low} we show the nucleus with the largest mass
fraction as a function of entropy and acceleration parameter.  As can be
seen from the figure, for high entropies Helium-4 has the largest mass
fraction.  The mass fractions of Helium-4 range from $X_{^{4}{\rm He}}=
0.30$ to $X_{^{4}{\rm He}}= 0.83$ in the region where it has a larger mass
fraction than any other nucleus. 

At high entropy there is a large number density of positrons which create 
protons in the outflow through the reaction 
$e^+ + n \rightarrow p +\bar{\nu}_e$.    
At low entropies, there are less positrons since the material is electron
degenerate, and therefore fewer protons. In addition for higher entropy,
nuclear statistical equilibrium favors light nuclei until relatively low
temperature.  Thus the triple alpha reaction freezes-out of equilibrium
before the majority of alpha particles can reassemble into heavy nuclei.
For low entropies, iron peak nuclei assemble at high temperature, before
the triple alpha reaction falls out of equilibrium. 

For low entropies we find that nucleus with the largest mass fraction is
$^{62}{\rm Ni}$.  The mass fraction in the region where $^{62}{\rm Ni}$
dominates ranges from $X_{^{62}{\rm Ni}}= 0.25$ to $X_{^{62}{\rm Ni}}=
0.75$. 

In the lower panel of Fig.\ref{fig:xmass_low} we show the nucleus with the
largest mass fraction {\it excluding Helium-4}. It is interesting to note
that when large amounts of Helium-4 are produced, Nickel-56 is produced as
well.  For more rapid accelerations heavier isotopes are produced and for
entropies of between $s/k = 30$ and $s/k = 40$ we see considerable
Zinc-66. 

\begin{figure}
\plotone{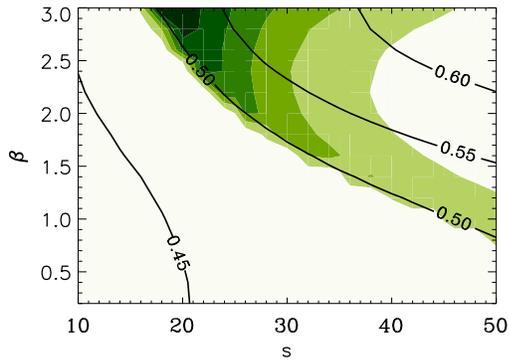}
\caption{Shows the mass fraction of $^{56}{\rm Ni}$ in the outflow of the low accretion
rate disk. Dark shaded regions correspond, from darkest to lightest, to
 mass fractions of greater than 0.5, 0.4, 0.3, 0.2, and 0.1. 
\label{fig:xmass_low_56ni}}
\end{figure}

Additional clues to the overall abundance pattern are given by the
electron fraction, $Y_e = 1/(1 + n/p)$, where $n/p$ is the neutron to
proton ratio. This is shown as the solid lines in Fig.
\ref{fig:xmass_low_56ni}. In addition to the effect of the entropy on the
electron fraction as already discussed, there is a significant change due
to the acceleration of the material.  Slowly accelerating material (high
$\beta$) has more time to approach the high equilibrium electron fraction
(as set by the entropy) than the rapidly accelerating material, which
retains some of the neutron richness of the disk.  In addition, with fast
acceleration (i.e. low $\beta$) there is little time for the neutrinos
flowing off of the disk to interact with the outflow material.  With slow
acceleration, however, there is more time and the electron fraction
increases. 

In this relatively low accretion rate disk the neutrinos are not trapped,
and therefore the neutrinos emitted from the disk are coming directly from
an inverse beta decay.  Since the disk is at low entropy, electrons in the
disk dominate over positrons. Therefore more electron neutrinos are
produced by inverse beta decay interactions than electron antineutrinos.
The electron neutrinos interact with the outflowing matter to raise the
electron fraction through the reaction $\nu_e + n \rightarrow p + e^-$. 
The effect of the neutrinos on the electron fraction here is relatively
small, of the order of  1\% - 5\%; however, this influence grows 
significantly in trajectories from disks with higher black hole spin 
parameter $a$ and higher accretion rates.

\begin{figure}
\plotone{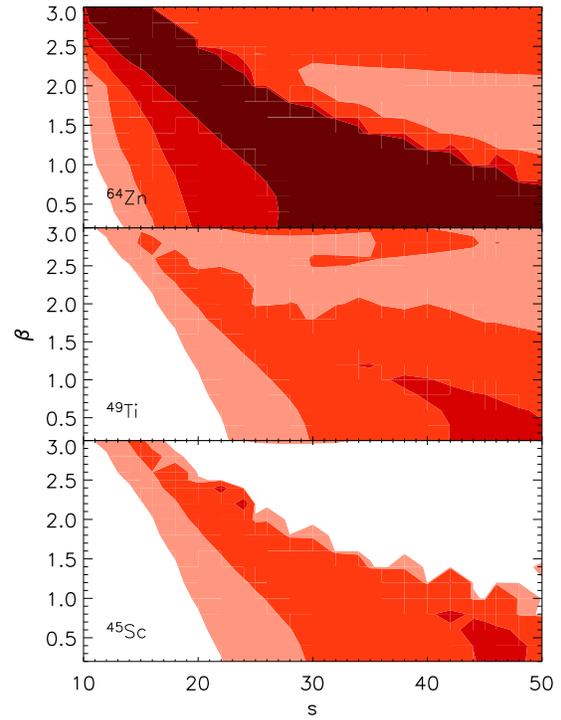}
\caption{Shows the overproduction factor of  $^{64}{\rm Zn}$, $^{49}{\rm Ti}$ and $^{45}{\rm Sc}$,
in the outflow of the low accretion
rate disk. Dark shaded regions correspond to overproduction factors, from
darkest to lightest, of greater than 1000, 100, 10, and 1. 
\label{fig:over_low_49ti}}
\end{figure}

In Fig. \ref{fig:xmass_low_56ni} the Nickel-56 abundance is shown.  The
largest abundance comes from moderate entropies and relatively slow
outflows.  \citet{psm04} suggested that for these trajectories which
produce considerable Nickel-56 an unusual nucleosynthesis pattern is
produced as well, including traditionally underproduced nuclei such as
45-Scandium, 49-Titanium, and 64-Zinc. Here we extend the results of that
study to include a larger range of trajectories that also include the
effects of neutrino scattering.  The results are shown in Fig.
\ref{fig:over_low_49ti}, which is a contour plot of the overproduction
factors for these nuclei, where we define overproduction factor as in
\citep{psm04}
\begin{equation}
O(j) = \frac{M_{wind}}{M_{SN ejecta}} \times 
\frac{X_{wind}}{X_{solar}},
\end{equation}
where $X_{solar}$ is the observed solar mass fraction of the nucleus,
$X_{wind}$ is our calculated mass fraction in the outflow, $M_{SN ejecta}$
is the mass ejected in a supernova explosion, and $M_{wind}$ is the total
mass contained in the outflow from the accretion disk.  Here we take
$M_{SN ejecta}$ to be 10 $M_{\sun}$ and $M_{wind}$ to be 1 $M_{\sun}$. All
three of these nuclei have large overproduction factors that extend in a
band from low entropy, slow acceleration to high entropy fast
acceleration.  The results correlate with electron fraction. 

\begin{figure}
\plotone{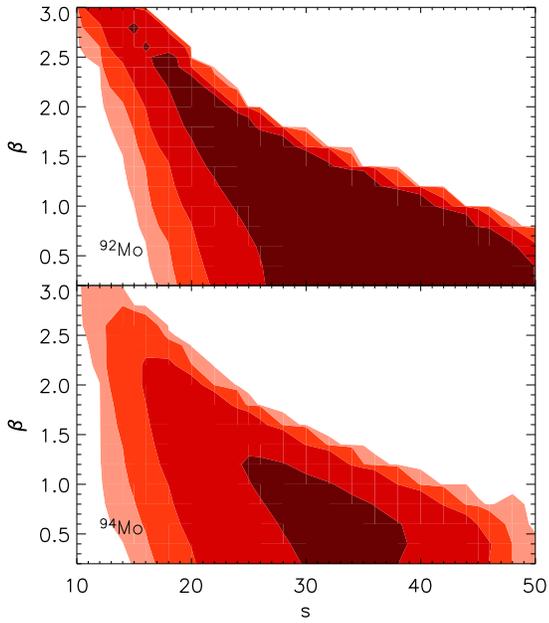}
\caption{Shows the overproduction factor of $^{92}{\rm Mo}$ and $^{94}{\rm Mo}$
 in the outflow of the low accretion
rate disk. Dark shaded regions correspond to overproduction factors,
in order from darkest to lightest of 1000, 100, 10, 1.
\label{fig:over_low_mo}}
\end{figure}

\begin{figure}
\plotone{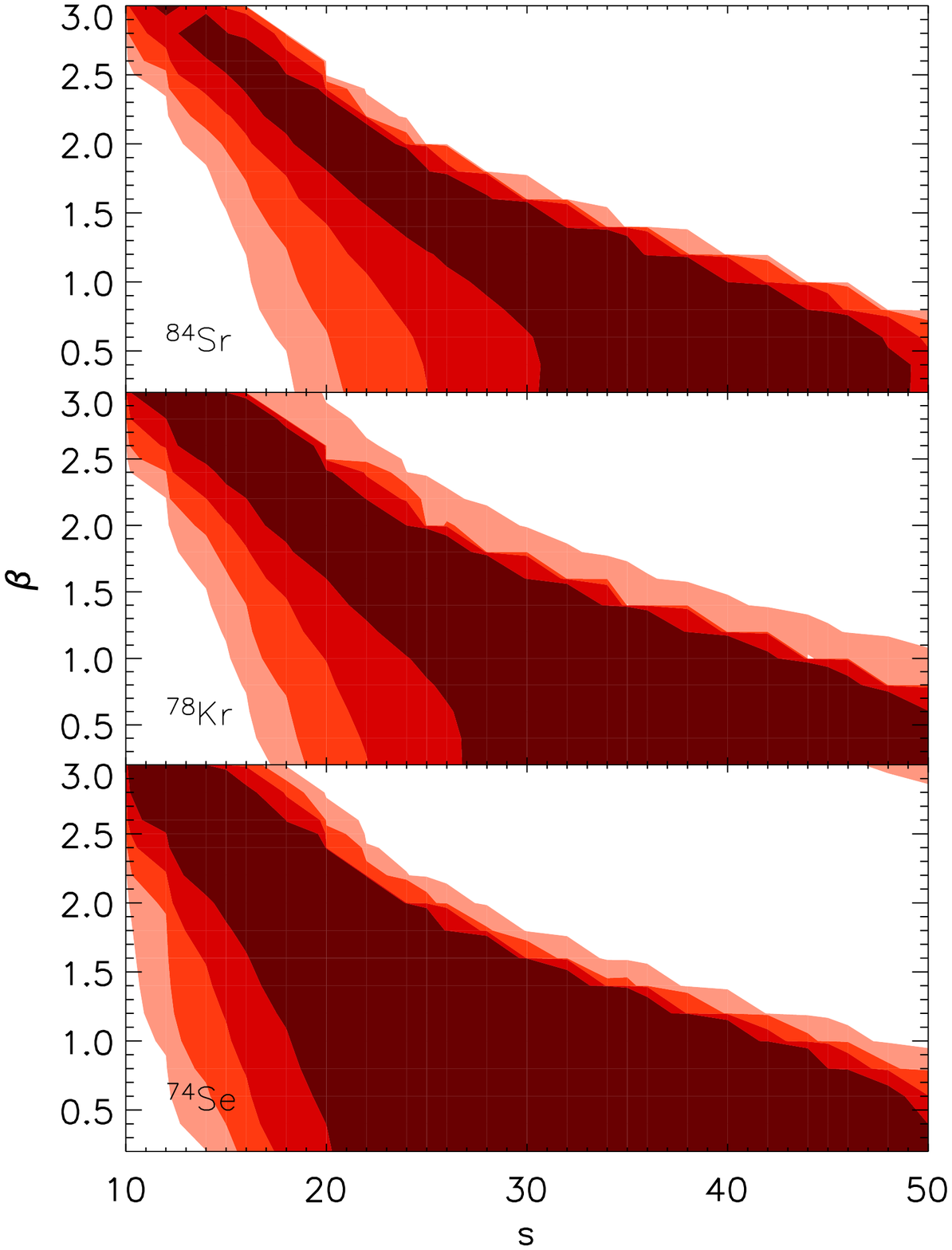}
\caption{Shows the overproduction factors of the
p-process nulcei  $^{84}{\rm Sr}$, $^{78}{\rm Kr}$,
and $^{74}{\rm Se}$
 in the outflow of the low accretion
rate disk. Dark shaded regions correspond to overproduction factors,
in order from darkest to lightest of 1000, 100, 10, 1.
\label{fig:over_low_sr}}
\end{figure}

In Fig. \ref{fig:over_low_mo} and \ref{fig:over_low_sr}
we consider also the light $p$-process nuclei, for which there is no
universally agreed upon production site.  It has been suggested in
\citet{fuji03} that these nuclei can be produced in an accretion disk.
We show Molybdinum-92 and Molybdinum-94 as well as
Strontium-84, Krypton-78 and Selenium-74 in the context of the outflow
from the low accretion rate disks.  We find overproduction factors for
$^{92}{\rm Mo}$ of greater than 1000 along the same band that overproduces
large amounts of $^{45}{\rm Sc}$, $^{49}{\rm Ti}$ and $^{64}{\rm Zn}$. 

The largest overproduction factors for Zinc occur for electron fractions
at just slightly less than one half. This is similar to the alpha rich
freeze-out of the neutrino driven wind of the supernova \citep{fm,hwmf},
but the overproductions we observe are larger. Overproduction factors
greater than 10 occur for conditions that are just slightly proton rich,
similar to what is observed in the hot bubble of the supernova
\citep{pruet,frohlich}.  There is another band of large overproduction
factor at slow outflow, which is due to the influence of neutrino 
capture on nucleons. It is interesting to note that in the observations 
of \citet{cayrel} the ratios [Zn/Fe] and [Zn/Mg] increase with decreasing 
metallicity.  While the origin of Zinc is a matter of debate, Magnesium 
is thought to be primarily produced in core collapse supernovae.

Other nuclei that have large overproduction factors in these disks include
$^{46}{\rm Ti}$,$^{59}{\rm Co}$, $^{88}{\rm Sr}$,$^{90}{\rm Zr}$,
$^{74}{\rm Se}$ and $^{63}{\rm Ni}$. 

\begin{figure}
\plotone{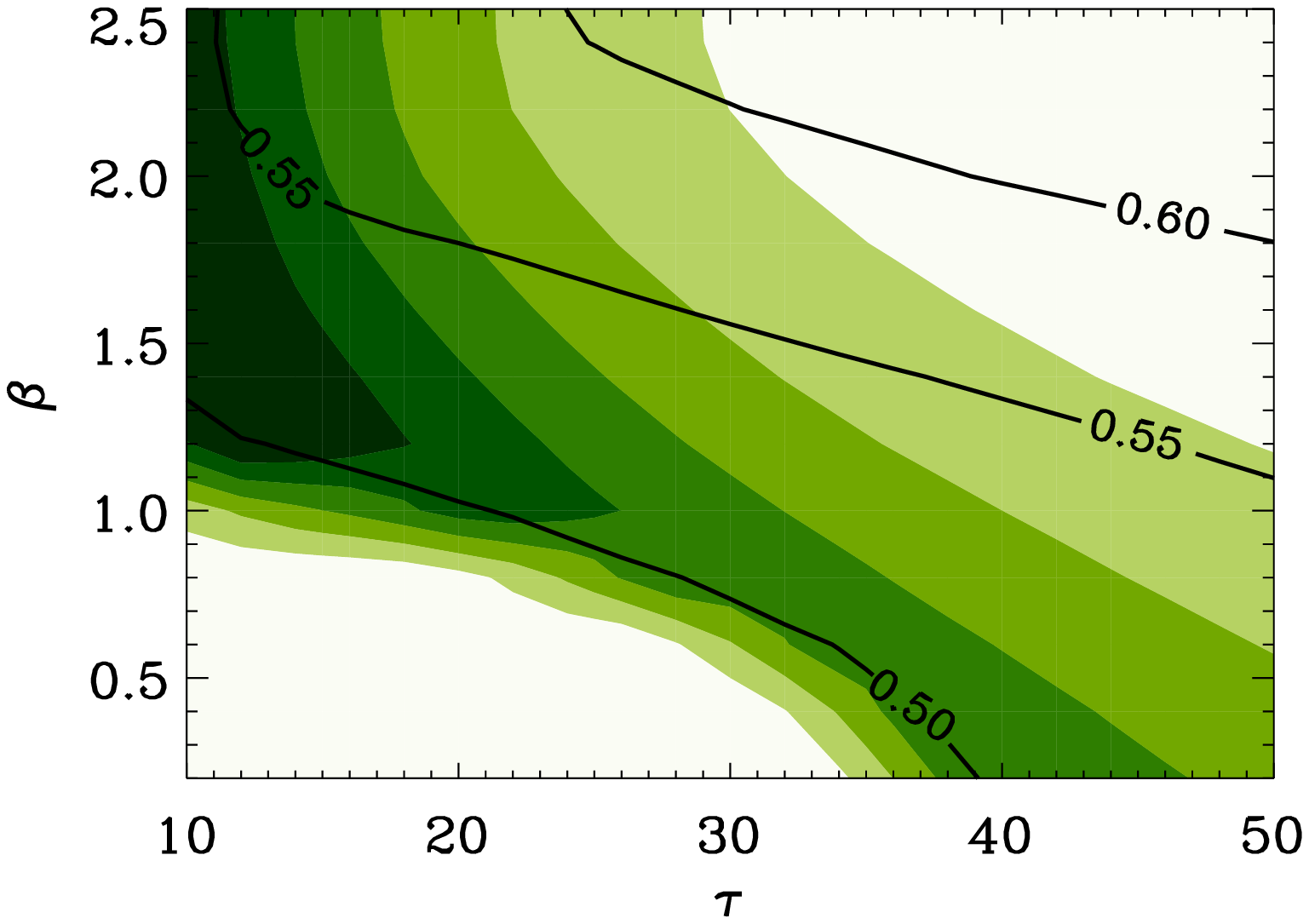}
\caption{Shows the mass fraction of $^{56}{\rm Ni}$ in the outflow of a 
low accretion rate disk, $\dot{M} = 0.1 \, 
{\rm M}_\odot / {\rm s}$, with black hole spin parameter $a=0.95$. 
Dark shaded regions correspond, from darkest to lightest, to mass 
fractions of greater than 0.5, 0.4, 0.3, 0.2, and 0.1. 
\label{fig:highspin}}
\end{figure}

As mentioned above the black hole spin parameter increases the degree
of trapping in the disk.  In Fig. \ref{fig:highspin} we show the
nickel-56 abundance for a PWF disk of $\dot{M} = 1 M_\odot$ / s and
$a=0.95$.  When compared with the electron fractions of the $a=0$ model shown
in Fig. \ref{fig:xmass_low_56ni}, it can be seen that the trapped neutrinos
significantly change the neutron-to-proton ratio.

\section{Element Synthesis from Medium Accretion Rate Disks} \label{medium}

Disks with accretion rates of $\dot{M} = 1 M_\odot$ / s are predicted by
simulations of neutron star merger events.  For these accretion disks the
electron neutrinos are trapped in the disk.  The antineutrinos are trapped
as well, but in a much smaller region than for the neutrinos. The number
of electron neutrinos overwhelms that of the electron antineutrinos coming
from the disk although the antineutrinos have higher temperature than the
neutrinos.  For the DPN model, the neutrino temperatures range from 4.1 
to 5.2 MeV and the antineutrino temperatures range from 5.0 to 
5.4 MeV \citep{sur04}.  Due to the overwhelming numbers of neutrinos, in the
outflowing material, the reaction $\nu_e + n \rightarrow p + e^-$ creates
very proton rich matter. 

For the lower accretion rate disks considered in the last section, even
the one with non-zero black hole spin parameter, the neutrinos make a
$\lesssim 30\%$ difference in the final value of the electron fraction,
but for the disk considered in this section the neutrinos dominate the
determination of the electron fraction and create a proton rich
environment in the disk outflow for trajectories with all accelerations
and entropies. 

\begin{figure}
\plotone{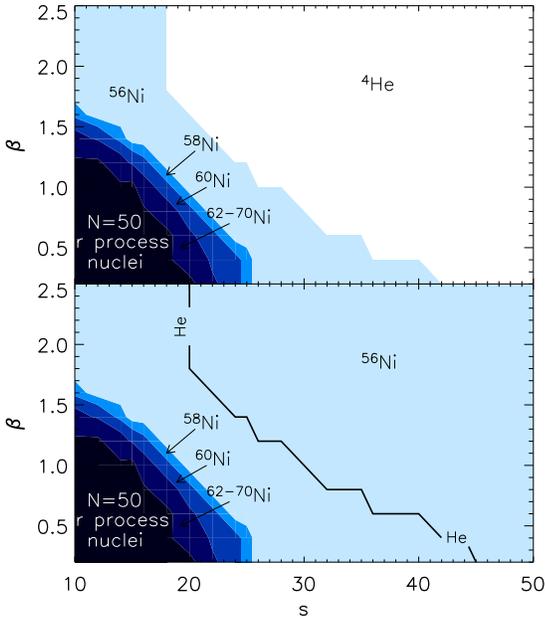}
\caption{Shows the elements with the largest mass fractions for the medium
accretion rate disk of $\dot{M} = 1 \, M_\odot / {\rm s}$.  The top panel
shows the element with the largest mass fraction and the bottom panel
shows the same {\it excluding He}.
\label{fig:xmass_med}}
\end{figure}

\begin{figure}
\plotone{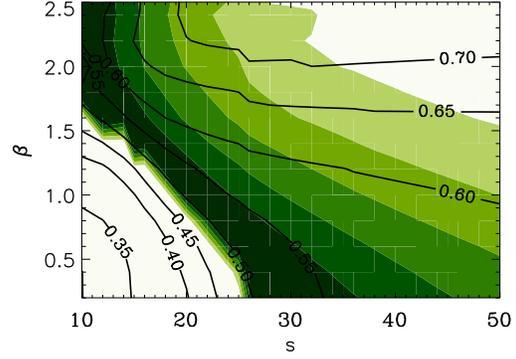}
\caption{Shows the mass fraction of $^{56}{\rm Ni}$ in the outflow of
 the low accretion
rate disk. Dark shaded regions correspond, in order from darkest to lightest, to
mass fractions of 0.5, 0.4, 0.3, 0.2, and 0.1.
\label{fig:xmass_med_56ni}}
\end{figure}

In the upper panel of Fig. \ref{fig:xmass_med} we show the nucleus with
the largest mass fraction for a range of accelerations and entropies. As
can be seen from the figure, helium again dominates the mass fraction for
larger entropies, with mass fractions of $\sim$0.3 - 0.6. In the lower
panel of Fig. \ref{fig:xmass_med} we show the nucleus with the largest
mass fraction excluding helium; again large amounts of $^{56}{\rm Ni}$ are
produced as well, with $X_{^{56}{\rm Ni}} \sim 0.2$ to $X_{^{56}{\rm Ni}}
\sim 0.7$ as shown in Fig. \ref{fig:xmass_med_56ni}. 

It can be seen from Fig. \ref{fig:xmass_med_56ni}, which shows the
electron fractions in the outflow that the material becomes very proton
rich, with $Y_e > 0.7$ for slow accelerations.  Given such proton rich
material it is interesting to investigate whether very light nuclei such
as Lithium and Boron are formed as well in the disk outflow. Table 1 shows
the abundances of these elements for a few sample trajectories.  For
comparison, observational values of $^7{\rm Li}/H$ at low metallicity are
in the range from $1 \times 10^{-10}$ to $2 \times 10^{-10}$ \citep{ryan}.
However, only a fraction of the baryons in the universe will pass 
through massive stars, so the subject bears further investigation with a 
galactic chemical evolution study.

It is generally thought that Boron-11, which has a solar mass fraction of
$X_{\odot,^{11}{\rm B}} \approx 5 \times 10^{-10}$, may be produced by a 
combination of cosmic ray spallation and the neutrino process, with 
estimates of up to 30\% for the latter contribution \cite{vangioni}. For 
a discussion of LiBeB from related site, hypernovae ejecta, see 
\citet{fields}.

\begin{deluxetable}{cccccc} 
\tablewidth{0pt}
\tablecaption{Light Element Synthesis in Outflows from $\dot{m}=1$ Disks}
\tablehead{
\colhead{S} & \colhead{$\beta$} & \colhead{$Y_{e}$} & 
\colhead{$X_{^{4}{\rm He}}$} & \colhead{$^{7}$Li/H} & \colhead{$^{11}$B/H}}
\startdata
10 & 2.5 & 0.560 & 0.102 & $2.2\times 10^{-9}$  & $6.7\times 10^{-10}$\\
15 & 2.5 & 0.641 & 0.191 & $9.6\times 10^{-10}$ & $1.6\times 10^{-10}$\\
30 & 2.5 & 0.767 & 0.314 & $1.6\times 10^{-10}$ & $1.1\times 10^{-11}$\\
40 & 2.5 & 0.779 & 0.372 & $3.7\times 10^{-11}$ & $2.0\times 10^{-12}$\\
\enddata
\end{deluxetable}

In Fig. \ref{fig:xmass_med_56ni} the mass fraction of $^{56}{\rm Ni}$ is
plotted. Most of the outflowing material becomes Nickel-56 for moderate
entropies $s/k \sim 30$, moderate outflows $\beta \sim 1.5$ or points in
between.  Little Nickel is produced fast outflows and low entropies as
well as for high entropies and slow outflows.  Our study suggests that if
collapsing stars form disks which accrete at a rate of $\dot{M} = 1
M_\odot / s$ or more slowly with higher spin parameter, then this range of
acceleration and entropy would be required to explain the nickel inferred
from the GRB light curve bumps. 

\section{Element Synthesis from High Accretion Rate Disks} \label{high}

\begin{figure}
\plotone{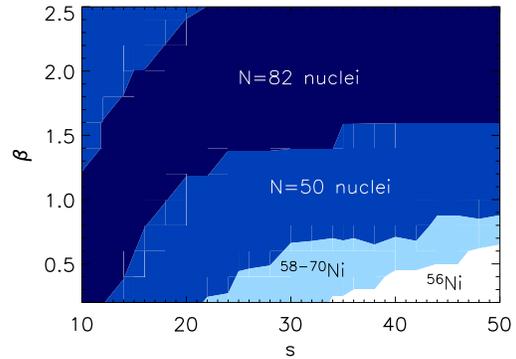}
\caption{Shows the elements with the largest mass fractions for the high
accretion rate disk, DPN $\dot{M} = 10 \, M_\odot / {\rm s}$. 
\label{fig:xmass_high}}
\end{figure}

In Fig. \ref{fig:xmass_high} we show the nucleus, or group of nuclei with
the largest mass fraction for all of the outflow trajectories from the
high accretion rate disk model, $\dot{M} = 10 \, M_\odot / \, {\rm s}$.
Rapid neutron capture process nuclei are produced abundantly for almost
all the trajectories considered, except for those with very rapid
acceleration and high entropy. The material in these trajectories becomes
very neutron rich because of the influence of the neutrinos. 

For such high accretion rates the electron antineutrinos are trapped in a
fairly large region, within a radius of 158 km from the black hole,
although the electron neutrinos are trapped in a larger region within
$r_{{\rm disk}}\sim 240$ km \citep{sur04}.  The temperature of the
electron antineutrinos is therefore higher, $3.6 < T_{\bar{\nu}_e} < 6.9 $
MeV as compared with $2.4 < T_\nu < 5.9$ MeV.  The neutrinos coming from
the surface of the disk interact with the outflowing matter. Because the
antineutrinos are more energetic than the neutrinos, the material is
driven neutron rich through the reaction $\bar{\nu}_e + p \rightarrow n +
e^+$ which has a larger flux averaged cross section than $\nu_e + n
\rightarrow p + e^-$. 

The degree of neutron richness depends on the neutrino fluence: the time
integrated neutrino flux experienced by a mass element in the outflow. 
For more slowly accelerating outflows the antineutrinos have more time to
interact creating more neutrons.  Conversely, for more rapid
accelerations, there is little time for neutrino interactions and the
electron fraction is higher. 

We note that for lower entropies, the material is electron degenerate with
very few positrons, and therefore has a large neutron-to-proton ratio 
regardless of its interactions with neutrinos or antineutrinos. 
We have also included the alpha effect as discussed in \cite{mmf}.

\begin{figure}
\plotone{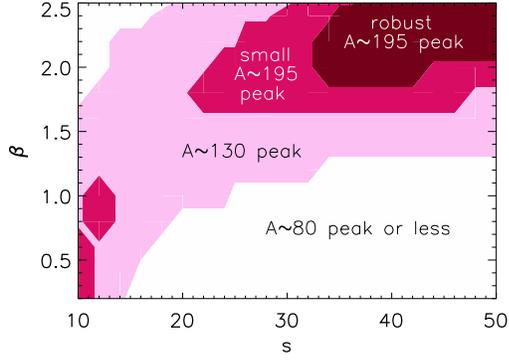}
\caption{For the high accretion rate disk, 
$\dot{M} = 10 \, M_\odot / {\rm s}$, shows the region where the 
second and third peaks of the $r$ process form.
 \label{fig:rprocess_cont}}
\end{figure}

Fig. \ref{fig:rprocess_cont} we show the peaks which are primarily formed
in the rapid neutron capture process.  We calculate the ratios between the
abundances in each peak region as in \citet{mb97}
\begin{eqnarray}
\frac{130}{80} & = &  \frac{\sum^{A=135}_{A=125} Y_{A}}{\sum^{A=75}_{A=85} Y_{A}} \cr
\frac{195}{130} & = & \frac{\sum^{A=200}_{A=190} Y_{A}}{\sum^{A=125}_{A=135} Y_{A}}
\end{eqnarray}

\begin{figure}
\plotone{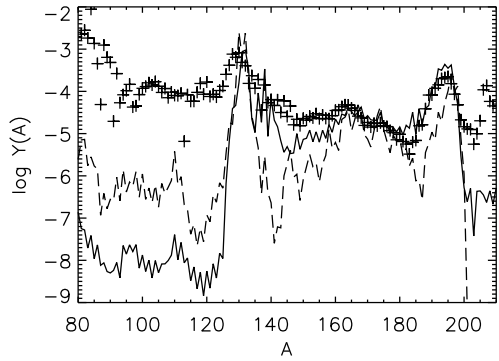}
\caption{Shows a final abundance pattern for outflow from the high
accretion rate disk with parameters $s/k=10$, $\beta=0.2$ (dashed line)
and $s/k=50$, $\beta=2.2$ (solid line).  Scaled solar system $r$-process
abundances are shown for comparison (crosses).
\label{fig:rprocess_abund}}
\end{figure}

The unshaded region is where the ratio 130/80 is less than 0.07; only the
$A=80$ peak is formed for these trajectories.  The ratio 130/80 is greater
than 0.07 and the ratio 195/130 is less than 0.3 for trajectories in the
next region, labeled $A=130$ as it is this peak that primarily forms.  The
darkest shaded region is where the ratio 195/130 is greater than 0.3 and
thus the heaviest elements are formed.  In the intermediately shaded
region, the ratio 195/130 is close to but below 0.3; in these trajectories
the $A=195$ peak has formed, but the relative peak heights are not in
agreement with the observed solar $r$-process abundances. A robust
$r$ process occurs in the slow acceleration trajectories.  The abundance
pattern for two trajectories, one with fast outflow and one with low
entropy is shown in Fig. \ref{fig:rprocess_abund}. 

For high entropy, the second, $A=130$, $r$-process peak is formed without
the $A=195$ peak from somewhat more quickly accelerating trajectories. 
And for very fast accelerations we find only the $A=80$ peak is formed. 
It is interesting to note that both meteoritic data \cite{qian98} and
observations of metal poor halo stars, e.g. \citet{sneden}, suggest two
$r$-process sites.  One of these forms the nuclei above the second peak
and one of these forms the nuclei below.  The meteoritic data suggests
that the third peak is associated with a relatively frequently occurring
event, such as core collapse supernovae. Core collapse supernova have been
studied extensively as $r$-process sites beginning with
\cite{meyer,woo94}, although at present there is no completely
self-consistent model that produces these elements without invoking new
particle physics. The meteoritic data also hint that the low mass
$r$-process nuclei are associated with a lower frequency event, such as a
neutron star merger \citep{argast}. 

Since the abundance distribution of the high mass nuclei in low
metallicity halo star tracks well the solar system abundance, such an
event should operate early in the evolution of the universe. An explosion
of a massive star, such as in the collapsar model, would also operate
early in the evolution of the universe.  However, this model presently
predicts lower accretion rate disks than the ones from which we find
$r$-process being produced.  In order to use this model to produce
$r$-process nuclei, the disk would need to pass through a stage where it
is very rapidly accreting or has a high black hole spin parameter. 

\begin{figure}
\plotone{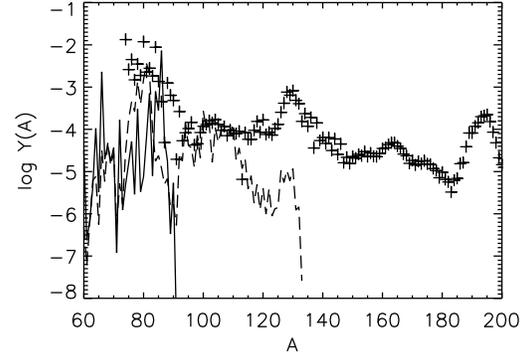}
\caption{Shows a final abundance pattern for outflow from the high
accretion rate disk with parameters $\beta=1.0$ and $s/k=30$ (dashed line)
and $s/k=45$ (solid line).  These trajectories give only the first peak of
the $r$ process. 
\label{fig:rprocess_abund_2}}
\end{figure}

The solar system abundances of lower mass $r$-process nuclei do not track
the abundance pattern measured in metal poor stars well.  This suggests
that the type of event that produces these nuclei does not necessarily
operate early in the evolution of the universe or is at least less
frequent. If the lower mass $r$-process nuclei were to come from disks
associated with neutron star mergers, then the outflow acceleration would
be fast $ \beta \sim 1$ for most possible values of entropy in the
outflow. In Fig. \ref{fig:rprocess_abund_2} we plot the abundance pattern
for two trajectories with entropies $s/k=30$ and $s/k=45$, both with
$\beta=1.0$. 

\section{Conclusions}

We find that a large range of nucleosynthesis may occur in the outflow
from gamma ray burst accretion disks.  For disks of $\dot{M} = 0.1 \,
M_\odot / s$ to $\dot{M} = 1 \, M_\odot / s$, most of the element
synthesis makes Helium and Nickel-56 except for low entropy and/or slow
acceleration trajectories. 

There is a significant, diagonal region of parameter space extending from
slow outflow, low entropy to fast outflow high entropy trajectories, where
nuclei such as $^{45}{\rm Sc}$, $^{49}{\rm Ti}$, $^{64}{\rm Zn}$,
$^{92}{\rm Mo}$, and $^{94}{\rm Mo}$ are overproduced by factors of
100-1000 in the low accretion rate models.  These low accretion rates are
typical of the collapsar model.  Disk winds can significantly contribute
to the galactic inventory of these nuclei if the rate of supernovae which
produce disks is $ > 10^{-2}$ times that of the rate of core collapse
supernovae.  This rate can be considerably higher than the limits on GRBs
of $\lesssim 10^{-2}$ since not all disks may produce bursts. 

In the outflow from both of these disks significant Nickel-56 is produced,
with mass fractions of up to 0.7.  Low accretion rate disks require slow
outflow and moderate entropy, $s/k \sim 20$, to synthesize considerable
Nickel-56, while moderate accretion rate disks can synthesize considerable
Nickel with faster outflow for similar entropies.  Introducing a non-zero
spin parameter to the disk creates more $^{56}{\rm Ni}$ at lower entropies
and outflow accelerations, due to the increased influence of the
neutrinos. 

The moderate accretion rate disks can also make the $N=50$ peak of the $r$
process for low entropy and fast outflow. The moderate accretion rate
disks are predicted to come from neutron star mergers.  Observational
evidence is consistent with such an infrequently occurring event which
synthesizes the $A=80$ peak. 

For the high accretion rate disks the majority of the parameter space
produces an $r$ process. The $A=195$ peak is synthesized only for slowly
accelerating outflows, although the $A=130$ peak is made more robustly. 
In the very fast acceleration, high entropy trajectories we find only the
first peak of the $r$ process. 

Future studies of the hydrodynamics of the outflow from accretion disks
surrounding black holes will determine which of these patterns of element
synthesis occurs in nature. 

\acknowledgments
We thank Jason Pruet for useful discussions.
We thank the Institute for Nuclear Theory at the
University of Washington for its Hospitality and the Department of 
Energy for partial support during the completion of this work. This
work was partially supported by the Department of Energy under
contract DE-FG02-02ER41216 (GCM) and the Research Corporation under
contract CC5994 (RS).

\end{document}